\documentclass{aastex}
\usepackage{emulateapj5}

\slugcomment{ Submitted, 5 April 2001; Revised, 29 April 2001; Accepted, 30 May 2001}

\shorttitle{High velocity white dwarfs}

\shortauthors{Reid {\it et al.}}

\begin {document}
\title{ High-velocity white dwarfs: thick disk, not dark matter}

\author {I. Neill Reid, Kailash C. Sahu}
\affil {Space Telescope Science Institute, 3700 San Martin Drive, Baltimore,
MD 21218; \\
e-mail: inr@stsci.edu, ksahu@stsci.edu}

\author {Suzanne L. Hawley}
\affil {Dept. of Astronomy, University of Washington, Seattle, WA 98195; \\
e-mail: slh@pillan.astro.washington.edu}

\begin{abstract}

We present an alternative interpretation of the nature of the extremely
cool, high-velocity white dwarfs identified by Oppenheimer 
{\sl et al.} (2001)
in a high-latitude astrometric survey. We argue that the velocities
of the majority of the sample
are more consistent with the high-velocity tail of a rotating
population, probably the thick disk, rather than a pressure-supported 
halo system.  Indeed, the observed numbers are well
matched by predictions based on the kinematics of a
complete sample of nearby  M dwarfs. 
Analysing only stars showing retrograde motion gives a local
density close to that expected for white dwarfs in the 
stellar ($R^{-3.5}$) halo. 
Under our interpretation, none of the white dwarfs need be assigned 
to the dark-matter, heavy halo.
However, luminosity functions derived from observations of these
stars can set
important constraints on the age of the oldest stars in the Galactic Disk.

\end{abstract}


\section {Introduction}

Tracking down the nature of dark matter could be described as the 
astronomical obsession of the twentieth century.
The concept emerged through Oort's comparison of the census
of luminous material (stars, gas, dust) in the Solar Neighbourhood
with dynamical estimates of the local mass density derived from the
motions of K giants, high in the Galactic Disk (Oort, 1932). Like
Zwicky's (1933) near-contemporary suggestion that cluster-galaxy kinematics
required dark mass on much larger scales, the problem of the 
local `missing mass' was not taken up immediately. However,
the discrepancy in mass densities, a factor of $\sim3$, prompted a
flurry of survey activity in the 1960s and 70s, largely centred on
what proved to be substantial overestimates of the number density of M dwarfs 
(see Reid \& Hawley, 2000, ch. 7). Debate continues
over the significance of the Oort limit discrepancy (Bahcall 
{\sl et al.}, 1992; Cr\'ez\'e {\sl et al.}, 1998).

In a cosmological context, dark matter came to prominence with the realisation that
rotation curves of many galaxies were not Keplerian at radii beyond the visible extent of
the disk (e.g. Rubin, Ford \& Thonnard, 1978). Some were flat,  
implying that the enclosed mass increases linearly with radius (Ostriker,
Peebles \& Yahil, 1974). Extrapolating to the Holmberg
radius, the total mass approaches $\sim10^{12} M_\odot$, with
mass-to-light ratios more than an order of magnitude  higher than expected
for an old stellar population. Ostriker {\sl et al.} suggested that the
additional material could be distributed in a near-spherical 
structure (a `heavy halo'), thereby dynamically stabilising the disk 
in spiral galaxies
(Ostriker \& Peebles, 1973). The implied radial density law is
$\rho_{HH}(R) \propto R^{-2}$, where $R$ is the distance from the centre of the Galaxy. 
This is substantially 
flatter than the $\rho_S(R) \propto R^{-3.5}$ density distribution of the 
$\sim3 \times 10^9 M_\odot$ traditional (Population II) stellar halo (Schmidt, 1975).

Subsequent developments in the field 
have been reviewed by Trimble (1987), Fich \&
Tremaine (1991) and Ashman (1992). The most
recent estimates, based on satellite
galaxy motions, place the mass of the Milky Way at
$\sim5 \times 10^{11} M_\odot$ for $R < 50$ kpc. (Wilkinson \& Evans, 1999).
Approximately 10\% of this mass can be accounted for by stars, gas and dust
in the Galactic disk and the stellar halo (Alcock {\sl et al.}, 2000).
Dark matter candidates for the remaining 90\% include exotic particles,  cold
molecular gas, and compact objects, ranging from 10$^6 M_\odot$ black holes
through brown dwarfs to space rocks. Of these categories, the
last is potentially most accessible to direct astronomical observation
through gravitational microlensing (Paczynski, 1986). A number of
extended photometric monitoring campaigns have been conducted, 
directed towards the high star-density regions of the LMC, SMC
and the Galactic Bulge, and a substantial number of lensing events detected. 

The contribution made by MACHOS to the dark halo 
is still uncertain. From their most recent analysis
of data taken in the central regions of the LMC, the MACHO group have
detected 17-19 events, from which they estimate the contribution of
MACHOs to dark matter to be about 20\%, $\sim 9 \times 10^{10} M_\odot$
(Alcock et al. 2000).  On the other hand, data from
the EROS group cover a wider region of the LMC, within which they have
detected at most 2 events. Based on those detections, 
they estimate that MACHOS contribute between 
0 and 20\% of the heavy halo (Lassere et al. 2000). While the exact
locations of the lenses remain controversial, mainly because the simple
microlensing light curve is inadequate to provide information  on the
lens locations, there are two binary events -- MACHO 98-SMC-1 and MACHO
LMC-9 -- for which the lens locations have been determined with
reasonable certainty (Albrow et al. 1999, Alcock et al. 1997, 1999).
In both these cases, the caustic crossing
timescales suggest that the  lenses are most likely in the
Magellanic Clouds, and this has been used to argue that most of the
lenses may be Magellanic Cloud members (e.g. Graff, 2000; Sahu and
Sahu, 1998). Furthermore, the very different time scales of the events
observed towards the LMC and the SMC seem to be inconsistent with 
MACHOs located in the halo (Sahu, 1994; 2001).

Whether MACHOs are Galactic or members of the Magellanic Clouds, it 
is clear that they are a minority constituent of the heavy halo. Moreover, 
the best estimate of the mass of the foreground lensing objects, based on
the distribution of event timescales and the assumed kinematics, 
proves to be surprisingly high, $\langle M \rangle = 0.5\pm0.3 \ M_\odot$.
Indeed, probably the most significant result from the microlensing projects is
the stringent limit set on possible dark matter contributions from lower-mass
objects such as brown dwarfs. 

The set of compact astronomical objects with $M \sim 0.5 M_\odot$, and
luminosities sufficiently low to escape detection, is small; white
dwarfs are the obvious candidate, presumably fossil remnants of a
primordial stellar population (Population III).
Such an hypothesis has significant implications, particularly for star 
formation theory: the corresponding mass function, $\Psi(M)$,  must 
peak strongly at intermediate masses, avoiding over-enrichment of Population
II by supernovae, and matching the number of long-lived low-mass dwarfs
to the stringent constraints set by general starcounts (Graff \& Freese, 1996;
Reid {\sl  et al.}, 1996).  While there is only weak evidence for radical
changes in $\Psi(M)$ in the abundance range spanned by disk and halo, a number
of theoretical models have been devised which accomodate those requirements 
at primordial abundances (Chabrier {\sl  et al.}, 1996; Chabrier, 1999). Even those models, 
however, produce sufficient ejecta that they have difficulty matching 
the observed abundance ratios of light elements in Population II stars 
(Gibson \& Mould, 1998; Fields {\sl  et al.}, 2000).

Accepting these caveats, it is possible to search for the white
dwarf members of the purported Population III heavy halo using their
unusual colours. 
Recent theoretical models (Bergeron {\sl  et al.}, 1995; Hansen, 1999) predict
that old, cool,
hydrogen-atmosphere white dwarfs have flux distributions which depart
significantly from blackbody curves, due mainly to H$_2$ absorption 
(Mould \& Liebert, 1978).  A small number of examples have been
found in the field (e.g. LHS 3250, Harris {\sl et al.}, 1999), although
none have the low luminosities (M$_R \sim 17$) expected for 12-14 Gyr-old degenerates.
Such objects might have escaped attention in previous large-scale surveys.

Almost contemporaneously with the theoretical work, analysis by Ibata {\sl et al.} (1999)
of second-epoch images of the Hubble Deep Field, taken two years after
the first-epoch data, led to the tentative identification of two to
five very faint, blue, point-like sources which appeared to exhibit
measureable proper motion. If those sources were heavy-halo white dwarfs, the mass
density would be sufficient to account for all of the heavy-halo dark matter.
However, the proper motions have recently been withdrawn 
based on third epoch HST observations, taken three years 
after the second-epoch observations (Richer, 2001).

The local mass density of the heavy halo is 
$\sim10^{-2} M_\odot \ {\rm pc}^{-3}$ (for an $R^{-2}$ mass dependence).
If 0.5$M_\odot$ objects contribute 20\% of the dark halo, the Solar
Neighbourhood density is 0.004 MACHOs pc$^{-3}$. Ibata {\sl et al.}
(2000) identify two nearby cool white dwarfs, which they suggest are
potential examples of this type of object. Flynn {\sl et al.} (2001), however, 
failed to identify any likely counterparts amongst published proper
motion surveys, while Monet {\sl et al.} (2000) identified only
a single candidate halo white dwarf in
a proper motion survey combining accepted and
rejected POSS II plate material in 35 fields ($\sim1380$ square degrees).
Although the latter object is spectral type DC, the spectrophotometric properties
are consistent with a cooling-age significantly younger than 12 Gyrs, making
it likely to be a member of the stellar halo. Finally, Harris {\sl  et al.} (2001) 
report the detection of an extremely cool white dwarf, SDSSp J133739.40+000142.8,
from SDSS commissioning data. However, the system has a low space motion, and is 
probably a low-mass disk star with a helium core.

Most recently, Oppenheimer {\sl et al.} (2001; O2001) have completed a 
deep proper-motion survey towards the South Galactic Cap and have identified 
38 cool white dwarfs. Based on kinematics, they assign
those stars to a halo-like population, and argue that most are members of the
heavy halo. If confirmed, this would be a result of considerable significance. 
However, we argue that the O2001 analysis fails to meet the burden of
proof required for such a radical hypothesis.
We present an alternative scenario, explaining the observations in terms
of conventional stellar populations. In the following section
we discuss the Oppenheimer {\sl et al.} observations in the broader
context of stellar population kinematics; the final section
summarises our conclusions.

\section {Galactic populations and local kinematics}

\subsection {Heavy-halo white dwarfs?}

Oppenheimer {\sl et al.'s} sample is drawn from 196 Schmidt fields,
covering 4165 square degrees towards the South Galactic Cap. They derive 
BRI magnitudes and proper motions for their sample by combining 
measurements of IIIaJ, IIIaF and IVN plate material from the UK/AAO Schmidt.
The (B-R) colours are used to estimate photometric parallaxes from
a linear colour-magnitude relation, with uncertainties of $\sim20\%$.
Radial velocities have not been measured for any of the white dwarfs; indeed,
most have featureless DC-type spectra. However, towards the Galactic Poles,
transverse motion depends most strongly on the U and V 
velocity components, motion towards the Galactic Centre and 
in the direction of Galactic rotation respectively. Thus, while the
available data do not permit a reliable estimate of W, the velocity
perpendicular to the Galactic Plane, planar (U, V) velocities can
be estimated from the astrometric data. 

Oppenheimer {\sl et al.} derive U and V velocities for each target, 
allowing for Solar Motion ({\sl via} secular parallax), 
and setting W=0 kms$^{-1}$ for each star. Systems with 
\begin{displaymath}
\sqrt{U^2 + (V+35)^2} \ > \ 94 \ {\rm kms^{-1}}
\end{displaymath}
fall outwith the limits of a velocity ellipsoid defined by the 2$\sigma$
motions of disk stars (from Chiba \& Beers, 2000; CB), and 
are identified as having halo-like kinematics. Using the $\Sigma {1 \over V_{max}}$
method, and limiting the sample to R$_{59F} < 19.7$,
they derive a density of
\begin{displaymath}
\rho_{WD}({\rm HH}) \ = \ 2.2  \times 10^{-4} \ {\rm stars \ pc}^{-3}
\end{displaymath}
This is a factor of ten higher than the expected density of white dwarfs
in the stellar halo, and Oppenheimer {\sl et al.} argue that 
these degenerates are local representatives of dark matter in the 
Milky Way's heavy halo. We note that the faintest white dwarf is
WD0351-564, with M$_R \sim 15.9$, almost 1 magnitude brighter than predicted
for 12-14 Gyr-old degenerates (Hansen, 1999). 

We have computed motions for the O2001 dataset using a
slightly different assumption; we adopt the O2001 photometric distance estimate (and 
associated uncertainty) and proper motion measurements, but set V$_{rad} = 0$
 kms$^{-1}$ rather than setting W=0 kms$^{-1}$. 
Since we have no information on either W or V$_{rad}$, both assumptions are 
equally valid. Moreover, if the O2001 interpretation of these results is correct,
it should be sufficiently robust to survive either assumption. We note that the
mean W velocity for the sample under our assumption is $0.6\pm54$ kms$^{-1}$,
giving no indication of our having introduced a significant bias. We also
allow for the Solar Motion, adopting U$_\odot$=10.0 kms$^{-1}$,
V$_\odot$=5.3 kms$^{-1}$ and W$_\odot$=7.2 kms$^{-1}$, from Dehnen \&
Binney's (1998) analysis of Hipparcos data.

Figure 1 plots the resulting (U, V) distribution, where we also show CB-disk 1$\sigma$ and
2$\sigma$ ellipsoids used by O2001, and those appropriate 
to  a non-rotating halo population with
dispersion 120 kms$^{-1}$. The most striking  feature, in 
both Figure 1 and Oppenheimer {\sl et al.'s} Figure 3, is the concentration
of stars near the boundaries of the CB-disk 2$\sigma$ ellipsoid.  
Moreover, there is a significant excess of prograde rotators, 34 of 38 stars
in our Figure 1, 30 of 38 in Figure 3 of O2001. 
This is exactly the behaviour expected if a sizeable fraction of the stars
are drawn from the high-velocity tail of a rotating population, 
rather than a non-rotating, pressure-supported halo.

Oppenheimer {\sl et al.} argue that the observed distribution is 
skewed by difficulties in detecting white dwarfs with high proper motions. 
In particular, they note that the probability of detecting stars with
$\mu > 3$ arcsec yr$^{-1}$ is less than 10\%. Such objects, however, should
make only a limited contribution to this survey, since  
with an average distance of 73 parsecs, $\mu = 3$ arcsec yr$^{-1}$ corresponds to
a transverse motion of 1040 kms$^{-1}$. This is a factor of two higher than
the highest velocity plotted in Figure 1, and, indeed, 
exceeds the local escape velocity. 

We can assess completeness at lower motions by plotting the cumulative 
distribution of proper motions.  
In a proper-motion limited r\'egime, one expects the number
of stars $N(\mu > \mu_{lim})$ to vary with $\mu^{-3}$. Figure 2 shows that
the O2001 halo candidates are broadly consistent with that distribution.
Stars at 73 parsecs with mildly retrograde orbits (V$_{tan}$ between 200 
and 350 kms$^{-1}$) have $0.57 < \mu < 1.01$. 
Thus, it seems unlikely that significant numbers of missing
high-$\mu$ stars could account for the scarcity of white dwarfs with 
retrograde motion.

\subsection {The velocity distribution of nearby M dwarfs}

Is there an alternative explanation of the nature of these stars? As
noted above, the concentration at high rotational velocities clearly
suggests an origin in the disk. Indeed, Oppenheimer {\sl et al's} choice 
of 2$\sigma$ velocity ellipsoids to differentiate disk and halo should
lead to significant contamination - 2$\sigma$ selection applied to two
uncorrelated parameters excludes only 86\% of a sample. In fact, disk
contamination does not reach that level since Chiba \& Beers' kinematics
are derived from a sample of metal-poor disk dwarfs 
($\langle [Fe/H] \rangle \approx -0.6$), and are therefore characteristic
of an older, higher velocity-dispersion sub-population. Assessing the
likely proportion of high-velocity disk white dwarfs demands kinematics for
a volume-limited sample of disk stars; local M dwarfs can supply that demand. 

Reid, Hawley \& Gizis (1995; RHG) have used spectroscopy and astrometry of almost
2000 M dwarfs to construct a volume-limited sample of 514 M-dwarf systems with
$8 < M_V < 15$. All of those stars have space motions determined to
an accuracy of 10-20 kms$^{-1}$ in each co-ordinate, allowing analysis
of the the kinematics of the local disk. Comparisons with spectroscopic surveys
suggest that the data are 95\% complete, with the missing stars predominantly
at low velocities\footnote{With the addition of new astrometric data, mainly from
Hipparcos, approximately 7\% of the RHG sample fail to meet the initial absolute 
magnitude-dependent distance
limits. Excluding those stars does not affect our conclusions - none of
the high-velocity stars are eliminated - so, for internal consistency, our analysis 
is based on the original RHG dataset.}. We can use these data to derive an 
estimate of the
parameter $f_{HVD}$,  the fraction of stars in an unbiased sample of
the Galactic disk which have
velocities exceeding the CD-disk 2$\sigma$ ellipsoid.

The overall velocity distributions are matched poorly by the
single Gaussian velocity dispersions implicit in the Schwarzschild ellipsoid model, 
particularly the rotational component, V, where there is evidence for multiple
components (RHG, Figure 11).  The distributions in U and W are simpler; the
main body of data can be represented as the sum of two distinct Gaussian components,  
with dispersions of $\sigma_U$=[35, 52] and $\sigma_W$=[20, 32] kms$^{-1}$ respectively.
Even in this model, there are indications of departures from Gaussian distributions
at high velocities - the r\'egime sampled by O2001. 
RHG suggested that 15-20\% of local stars
might reside in the higher velocity-dispersion component, but more detailed analysis shows that
the latter stars comprise no more than 12\% of the total. Allowing for incompleteness
at low velocities, the likely ratio of low-$\sigma$:high-$\sigma$ stars is
$\approx90:10$. Thus, the high-velocity component has a local density comparable to that
inferred for the `thick disk' in recent star-count analyses (Ohja {\sl et al.}, 1999;
Siegel {\sl  et al.}, in prep.). The origins of that component, originally identified
by Gilmore \& Reid (1983), remain a subject of
debate, but, since velocity dispersion is a function of age (Wielen, 1977), the thick
disk is likely to include the oldest stars in the Galactic disk.

Given the complex kinematics shown by the M-dwarf sample, we do not attempt
to model the (U, V) velocity distribution. Instead, we conduct a simple empirical 
comparison. Figure 3 plots (U, V) data for the RHG M-dwarf sample.
Twenty systems lie beyond the 2$\sigma$ CB-disk contours,
and would be classed as candidate halo stars by Oppenheimer {\sl et al.} 
If we assume 95\% completeness in the RHG sample, this corresponds to a
fractional contribution of $f_{HVD} = 3.7\pm0.8$\%.
This fraction exceeds  the expected frequency of subdwarfs
in the Solar Neighbourhood, $\sim0.2\%$. 

Since the stars plotted in Figure 3 are M dwarfs, we can check whether 
the high-velocity objects have metallicities are consistent with 
halo membership. Population II
stars have [Fe/H]$<$-1; members of a Population III heavy halo should be 
significantly more metal-poor, with near-primordial abundances. 
M subdwarfs are recogniseable in having 
weaker TiO absorption than solar-abundance disk dwarfs with the same
CaH absorption (Gizis, 1997), but all of the outliers in Figure 3
have near-solar CaH/TiO bandstrength ratios (Figure 4). Thus, the high-velocity
M dwarfs are members of a disk population, probably the thick disk. 
It would be surprising if these main-sequence stars were to lack 
degenerate counterparts, and as the oldest disk stars, the cooling times
can be sufficiently long to achieve temperatures of less than 3000K, similar
halo white dwarfs.

\section {White dwarfs, dark matter and the thick disk}

 We have argued that the morphological similarities between Figures 1 and 3 indicate
that a sizeable fraction of the O2001 white dwarf 
sample are members of the Galactic disk.  Those similarities are emphasised 
in Figure 5, where  we superimpose the two
distributions. At least half of the O2001 white dwarfs are aligned in U with
the M-dwarf distribution (at 7 o'clock and 11 o'clock, relative to the 
centre of the disk ellipsoid).  

In order to estimate the likely number of high-velocity white dwarfs
contributed by the disk, 
we need to estimate the density of the parent population, 
comprising not only white dwarfs, but also their main-sequence progenitors.
The latter are included in the calculation since $f_{HVD}$
gives the fraction of high-velocity stars within a particular mass
range, summed over all ages. We assume that all stars, regardless of
mass, undergo the same statistical dynamical evolution over the
history of the Galactic disk. White dwarfs are the older members of
the subset of stars with masses in the range $\sim1$ to 8 M$_\odot$, just as
chromospherically inactive dM dwarfs are the older members of the
$\sim0.2$ to 0.5 M$_\odot$ stars plotted in Figure 2. The 
main-sequence progenitors need to be included in the intermediate-mass analysis, 
just as  dMe dwarfs are included in the sample plotted in Figure 2.
Thus, the expected density of high-velocity disk white dwarfs is given by
\begin{displaymath}
\rho_{HVD} \ = \ f_{HVD} \ . \ (\rho_{WD} + \rho_{MS})
\end{displaymath}

The simplest method of making this calculation is to use
statistics for nearby stars. 
Twelve white dwarfs within 8 parsecs are currently known: seven
single stars, two wide companions of low-mass red dwarfs (40 Eri B and
Stein 2051B), two companions
of massive stars (Sirius  and Procyon) and one unresolved companion
of an M-dwarf (G203-47B; see Reid \& Hawley, 2000). Excluding the binaries, the
local number density is 
\begin{displaymath}
\rho_{WD}({\rm disk}) \ = \ (3.26 \pm 1.23) \times 10^{-3} \ {\rm stars \
pc}^{-3}
\end{displaymath}
where the cited uncertainty reflects only Poisson statistics.
In comparison, Liebert {\sl  et al.} (1988) derive 
\begin{displaymath}
\rho_{WD}({\rm LDM}) \ = \ 3.0  \times 10^{-3} \ {\rm stars \ pc}^{-3}
\end{displaymath}
from analysis of white dwarfs found in proper motion surveys. A recent study
by Mendez \& Ruiz (2001), based on a more extensive sample, gives
\begin{displaymath}
\rho_{WD}({\rm LDM}) \ = \ 2.5 \times 10^{-3} \ {\rm stars \ pc}^{-3}
\end{displaymath}
Both are broadly consistent with the density estimate derived from the nearest stars.

The 8-parsec sample includes ten main-sequence stars (including Sirius and Procyon)
with masses exceeding 1 M$_\odot$ and main-sequence lifetimes less than 10 Gyrs, 
giving an effective local number density of 
\begin{displaymath}
\rho_{WD} + \rho_{MS}   = (7.9 \pm 1.9) \times  10^{-3} \ {\rm stars \ pc}^{-3}
\end{displaymath}
Applying the appropriate correction factor of $f_{HVD}=3.7\%$, we have
\begin{displaymath}
\rho_{HVD}({\rm disk}) \ = \ (2.9 \pm 0.7) \times 10^{-4}  \ {\rm stars \ pc}^{-3}
\end{displaymath}
Again, the uncertainty reflects counting statistics. 
White dwarfs in the stellar halo are also expected in the Solar Neighbourhood.
Assuming that the initial mass
function, $\Psi(M)$, is similar to that of the disk, as suggested by observations
of globular clusters (Piotto \& Zoccali, 1999), then these stars contribute an  
additional $2 \times 10^{-5}$ stars pc$^{-3}$ (Gould {\sl et al.}, 1998). 
Thus, based on the properties of the known stellar populations in the Solar
Neighbourhood, 
we expect a total density of high-velocity white dwarfs of
\begin{displaymath}
\rho_{WD}({\rm HV)} \ \approx \ (3.1 \pm 0.7) \times 10^{-4} \ {\rm stars \
pc}^{-3}
\end{displaymath}
This is entirely consistent with $\rho_{WD}({\rm HH)}$, the space density of
high-velocity white dwarfs derived by Oppenheimer {\sl  et al.} for their sample
of 38 cool white dwarfs. 

These calculations can be taken a step further. While the majority of
the white dwarfs in Figure 1 can be accomodated in the Galactic disk, 
a minority have substantial heliocentric velocities and are 
more likely to originate in a pressure-supported halo population.
In a non-rotating system we expect equal numbers of prograde and
retrograde rotators, so we 
can use the number of retrograde-rotating white dwarfs
to estimate the local density of that population. 
We measure V$< -220$ kms$^{-1}$ for four white dwarfs: F351-50,
LHS 147, WD0135-039  and WD0300-044. All
four are brighter than the R59F=19.7 completeness limit cited by Oppenheimer
{\sl et al.} for their survey. Summing ${1 \over V_{max}}$, allowing for 10\% sky coverage, 
gives a density of $\rho = 0.91 \times 10^{-5}$ stars pc$^{-1}$. Doubling that value to
take prograde rotators into account gives
\begin{displaymath}
\rho_{WD}({\rm halo}) \ = \ 1.8 \times 10^{-5} {\rm \ stars \ pc^{-3}}
\end{displaymath}
If we adopt the O2001 prescription for deriving (U, V), four additional white
dwarfs have retrograde motion: WD0153-014, LHS 542, WD0351-564 and LP586-51.
Including the contribution of the three stars which satisfy the
completeness limit (WD0351-564 has R59F=19.72) increases the  
derived density by less than 40\% giving 
\begin{displaymath}
\rho_{WD}({\rm halo}) \ = \ 2.4 \times 10^{-5} {\rm \ stars \ pc^{-3}}
\end{displaymath}
Either value is  consistent with that expected for
 white dwarfs in the Population II halo (Gould {\sl et al.}, 1998).

In summary, high-velocity disk stars can account for $\sim75\%$ of the
white dwarfs discovered by Oppenheimer {\sl et al.} (2001). The
remaining stars, particularly the retrograde rotators, 
are explained as members of the classical
halo. Given these results, we do not believe it
necessary to invoke any contribution from hypothetical
white dwarf members of a dark-matter, heavy halo.

Despite our disagreement over the interpretation of these
results, it is clear that the stars discovered 
in Oppenheimer {\sl et al.'s} South Galactic Cap survey represent a
significant addition to the catalogue of cool white dwarfs.
Besides providing further insight on cooling processes during
the final stages of evolution of degenerate dwarfs, statistical
analysis of the sample can provide important information on the
star formation history of the Galaxy. In particular, if we are
correct in our conjecture that the bulk of these stars are 
members of the thick
disk, then their luminosity function can be used to probe the age
of that sub-population, calibrating the first epoch of star formation
within the Galactic disk. 

\acknowledgements 
The authors thank Ben Oppenheimer for providing additional data for the
white dwarfs. We also thank Neal Dalal and Geza Gyuk for their comments,  
notably pointing out the difference
between one-dimensional and two-dimensional statistics.

\newpage

\begin{figure}
\plotone{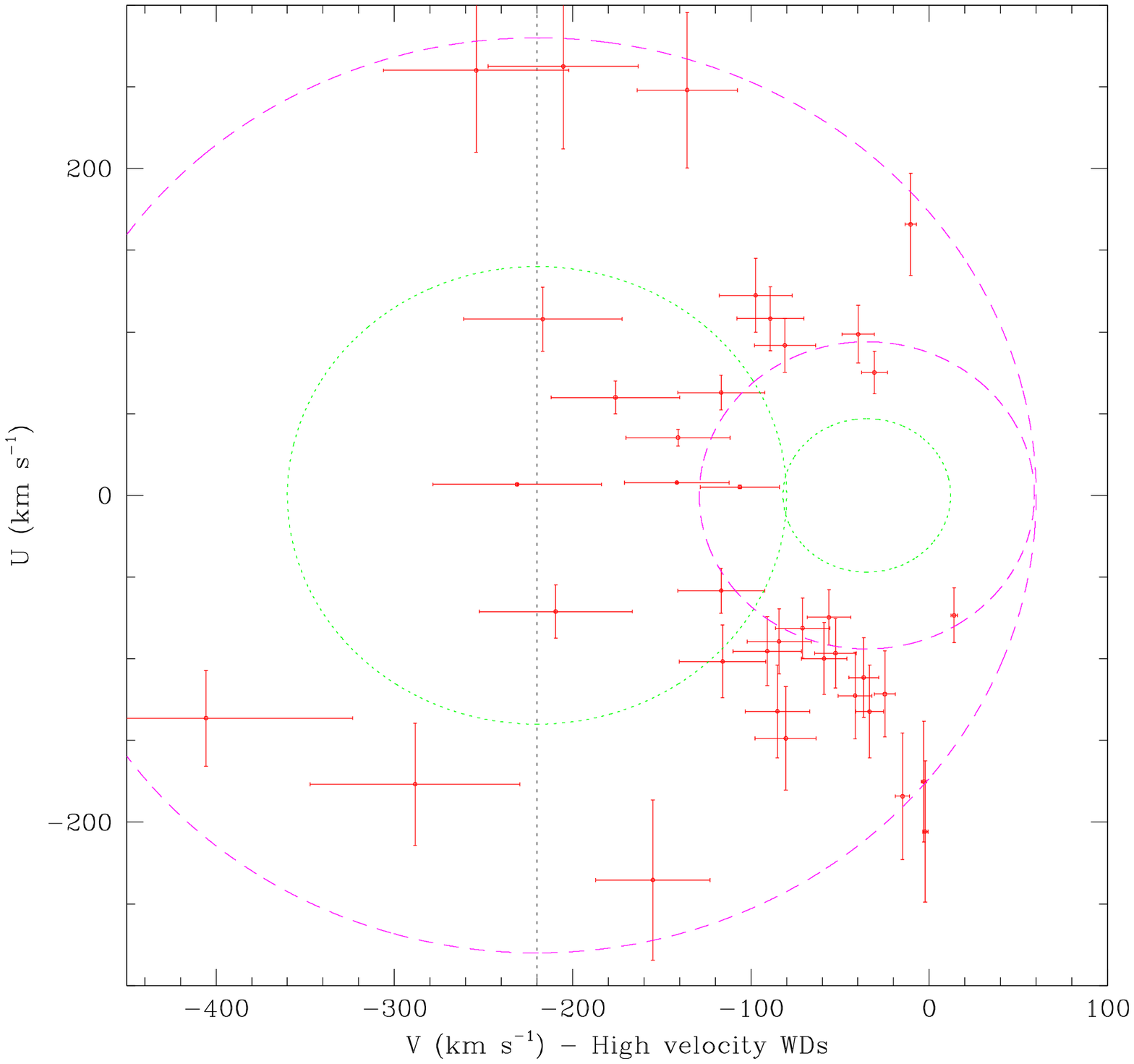}
\caption{The (U, V) velocity distribution for the 38 candidate halo
white dwarfs from Oppenheimer {\sl et al.}. The velocity zeropoint in V
is defined by the Local Standard of Rest; the vertical dotted
line marks V=-220 kms$^{-1}$, corresponding to zero net rotation. 
The ellipsoids plot the 1$\sigma$ (dotted) and 2$\sigma$ (dashed) 
contours for the Galactic disk and stellar halo populations, based on
kinematics from Chiba \& Beers (2000). The errorbars reflect distance
uncertainties of 20\%, and are therefore correlated in U and V. 
Other possible systematics are not taken into account.}
\end{figure}

\begin{figure}
\plotone{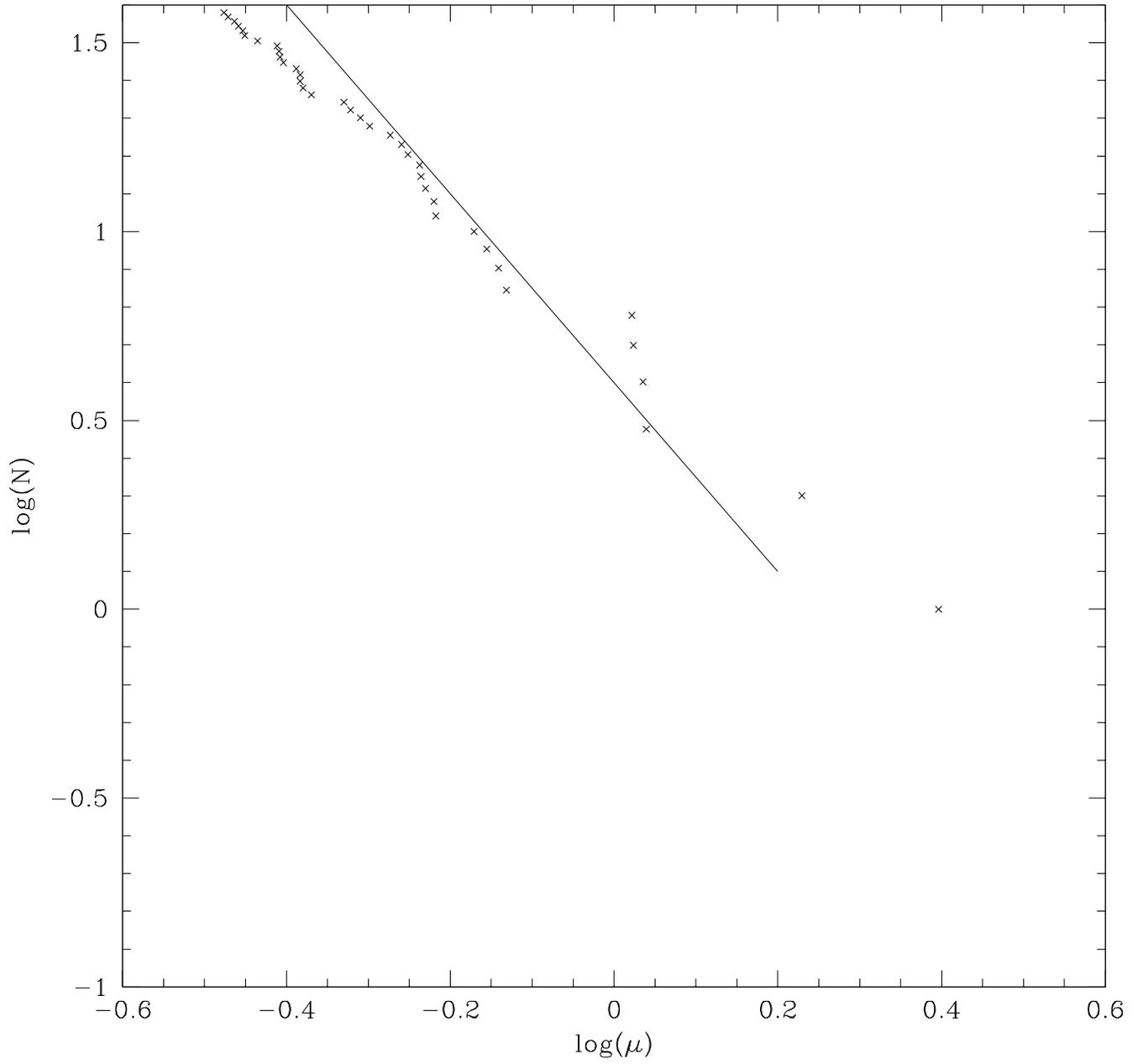}
\caption { The cumulative distribution of proper motions for the O2001 white
dwarfs. The solid line plots $N \propto \mu^{-2.5}$, shallower than 
expected for a volume limited sample. The comparison suggests only minor
incompleteness at high $\mu$, even at motions exceeding 1 arcsec yr$^{-1}$.}
\end{figure}

\begin{figure}
\plotone{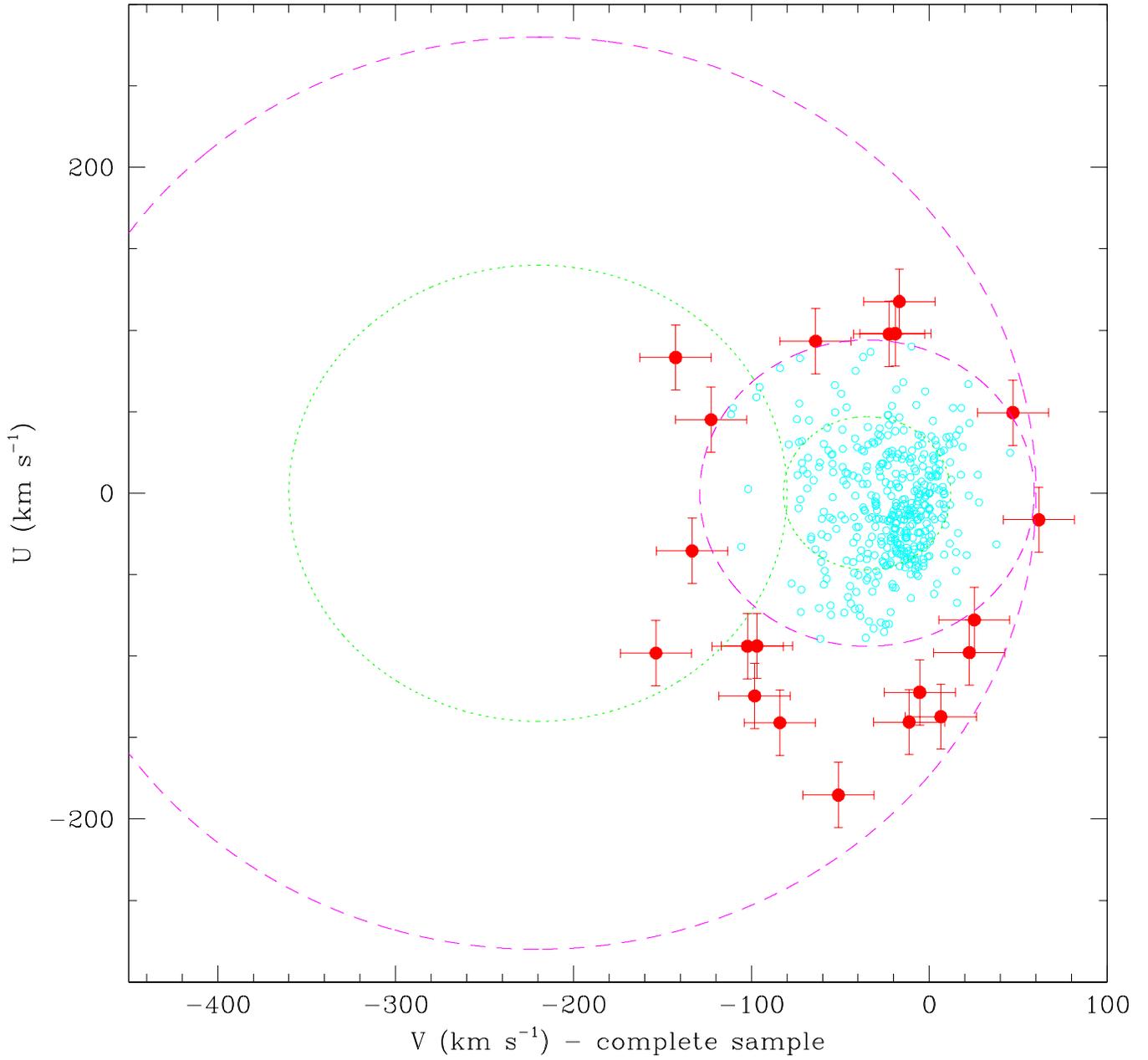}
\caption{The (U, V) velocity distribution for a volume-complete sample 
M-dwarf systems (from Reid, Hawley \& Gizis, 1995). All of the
stars are members of the Galactic disk;  20 stars, identified as
solid points with $\pm20$ kms$^{-1}$ errorbars, lie
outside the 2$\sigma$ disk velocity contours. }
\end{figure}

\begin{figure}
\plotone{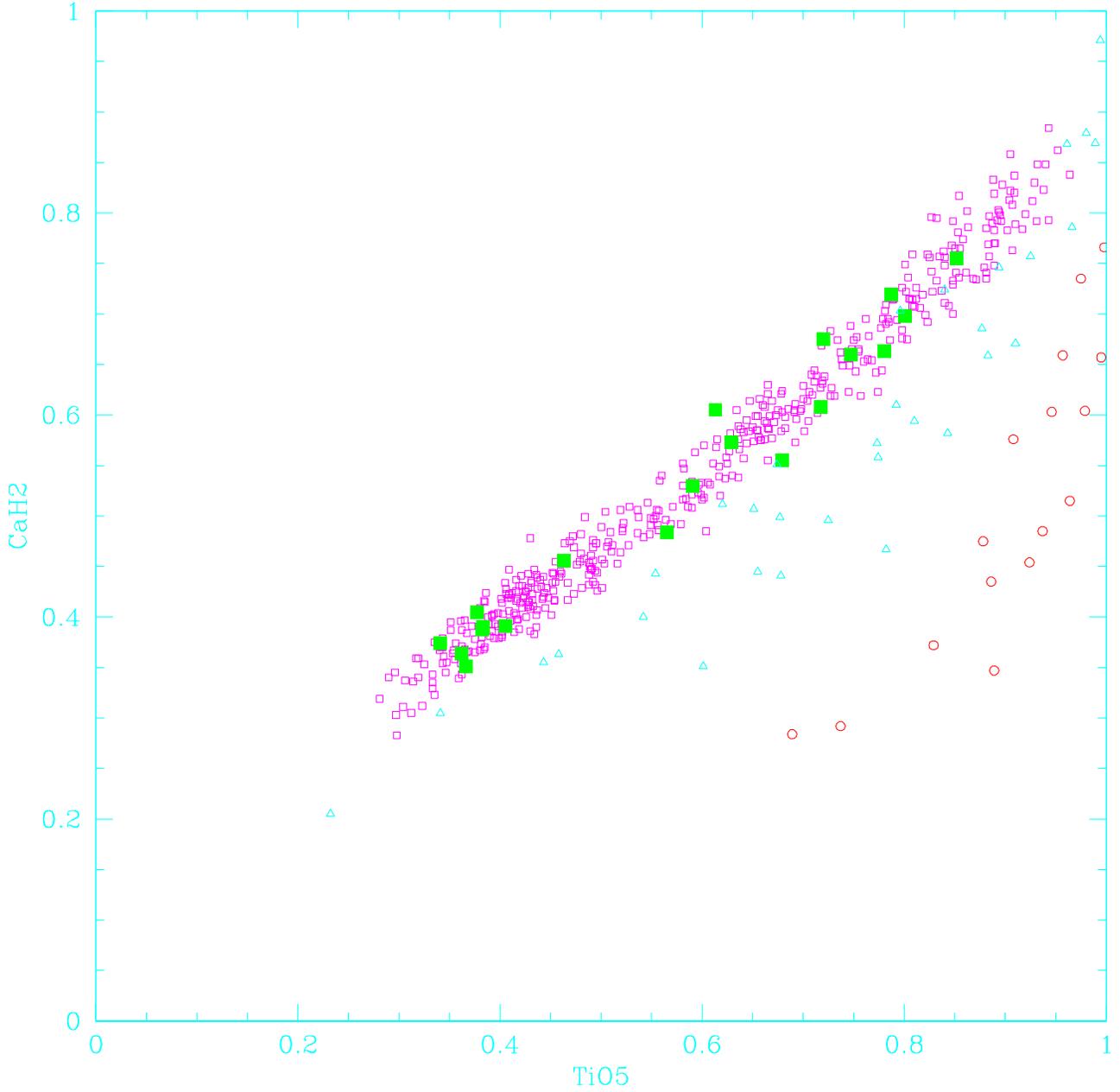}
\caption{The (TiO5, CaH2) diagram for late-type dwarfs: both TiO5 and CaH2
decrease in value with decreasing temperature, and the sequence runs from
$\sim$K7 to $\sim$M8. Data for the
complete sample of nearby dwarfs are shown as open squares, with solid
squares denoting the 20 high-velocity dwarfs. Metal-poor stars fall below
the disk sequence: intermediate-abundance
halo subdwarfs ([Fe/H]$\approx -1$) are plotted as open triangles,
while extreme metal-poor subdwarfs ([Fe/H]$\approx -2$) are plotted
as open circles (from Gizis, 1997). It is clear that the high-velocity
dwarfs have near-solar abundance.}
\end{figure}

\begin{figure}
\plotone{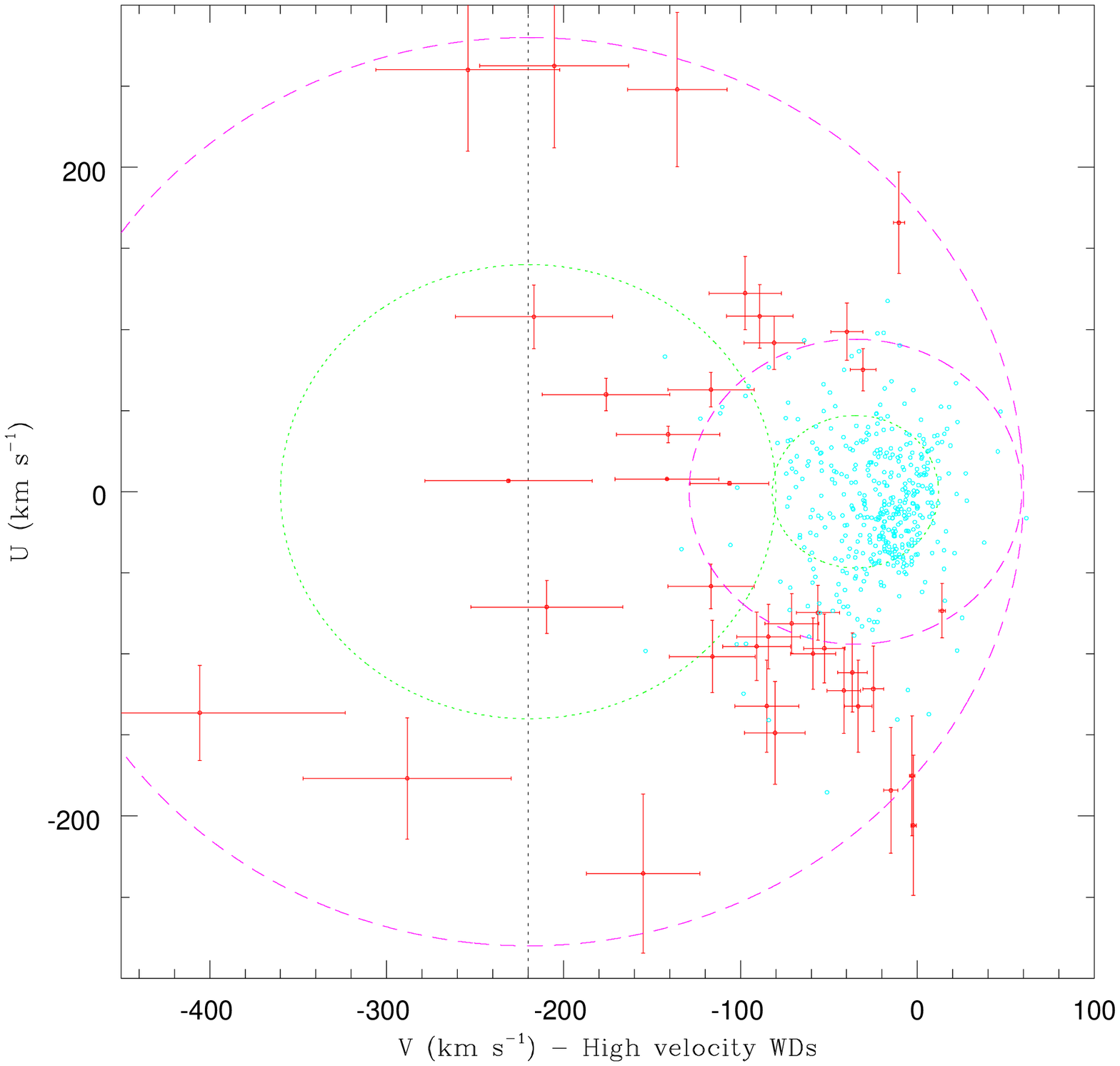}
\caption{The combined (U, V) velocity distribution of nearby M dwarfs (plotted as
open circles) and O2001 white dwarfs (points with error bars). Note the concentration
of the latter stars towards the U-velocity extremes of the M-dwarf distribution. The
contours have the same meaning as in Figures 1 and 3.}
\end{figure}

\newpage


\begin{references}
\overfullrule=0pt

\reference {} Albrow, M. {\sl et al.} 1999, ApJ 512, 672

\reference {} Alcock, C. {\sl et al.} 1997, ApJ 486, 697

\reference {} Alcock, C. {\sl et al.} 1999, ApJ 518, 44

\reference {} Alcock, C. {\sl et al.} 2000, ApJ 542, 281

\reference {} Ashman, K.M. 1992, PASP 104, 1109

\reference {} Bahcall, J.N., Flynn, C., Gould, A. 1992, ApJ 389, 234

\reference {} Bergeron, P., Saumon, D., Wesemael, F. 1995, ApJ 443, 764

\reference {} Chabrier, G., segretain, L., Mera, D. 1996, \apj 468, L21

\reference {} Chabrier, G. 1999 \apj 513, L103

\reference {} Chiba, M., Beers, T.C. 2000, AJ 119, 2843

\reference {} Cr\'ez\'e, M., Chereul, E., Bienaym\'e, O., Pichon, C. 1998,
A\&A 329, 920

\reference {} Dehnen, W., Binney, J.J. 1998, MNRAS 298, 387

\reference {} Fich, M., Tremaine, S. 1991, ARAA 29, 409

\reference {} Fields, B.D., Freese, K., Graff, D.S. 2000, \apj 534, 265

\reference{} Flynn, C., Sommer-Larsen, J., Fuchs, B., Graff, D. S., 
Salim, S. 2001, MNRAS 322, 553

\reference {} Gibson, B.K., Mould, J.R. 1997, ApJ 482, 98

\reference {} Gilmore, G.F., Reid, I.N. 1983, MNRAS 202, 1025

\reference {} Gizis, J.E. 1997, AJ 113, 806

\reference {} Gould, A., Flynn, C., Bahcall, J.N. 1998, ApJ 503, 798

\reference {} Graff, D. 2000, Microlensing 2000: A New Era of Microlensing
Astrophysics,  ASP conference proceedings, eds. J.W. Menzies and P.D.
Sackett  (Astro-ph/0005521)

\reference {} Hansen, B.M.S. 1999, ApJ 520, 680

\reference {} Harris, H.C., Dahn, C.C., Vrba, F.J., Henden, A.A.,
Liebert, J., Schmidt, G.D., Reid, I.N. 1999, ApJ 524, 1000

\reference {} Ibata, R., Richer, H., Gilliland, R., Scott, D. 1999, 
ApJ 524, L95 

\reference {} Ibata, R., Irwin, M., Bienaym\'e, O., Scholz, R., 
Guibert, J. 2000, ApJ 532, L41 

\reference {} Lassere, T. {\sl et al.} (the EROS collaboration), 2000, Astron. Astrophys.
355, L39 

\reference {} Liebert, J., Dahn, C.C., Monet, D.G. 1988, \apj 332, 891

\reference {} Mendez, R.A., Ruiz, M.T. 2001, \apj 547, 252

\reference {} Monet, D.G., Fisher, M.D., Liebert, J., Canzian, B., 
Harris, H.C., Reid, I.N. 2000, AJ 120, 1541

\reference {} Mould, J.R., Liebert, J. 1978, ApJ 226, L29

\reference {} Ohja, D.K., Bienaym\'e, O., Mohan, V., Robin, A.C. 1999, A\&A 351, 954

\reference {} Oort, J.H. 1932 BAN 6, 249

\reference {} Oort, J.H. 1960 BAN 15, 45

\reference {} Oppenheimer, B.R., Hambly, N.C., Digby, A.P., Hodgkin, S.T.,
Saumon, D. 2001, Science  292, 698

\reference {} Ostriker, J.P., Peebles, P.J.E., Yahil, A. 1974, ApJ 193, L1

\reference {} Paczynski, B. 1986, ApJ 304, 1

\reference {} Piotto, G., Zoccali, M. 1999, A\&A 345, 485

\reference {} Reid, I.N., Yan, L., Majewski, S.R., Thompson, I., Smail, I.  1996,   
\aj 112, 147

\reference {} Reid, I.N., Hawley, S.L., Gizis, J.E. 1995, AJ 110, 1838

\reference {} Reid, I.N, Hawley, S.L. 2000 {\sl New Light on Dark Stars},
(Praxis Publishing, Springer: London, Berlin)

\reference {} Richer, H. 2001, in {\sl The Dark Universe:  Matter, Energy, and Gravity},
STScI, (in prep.)

\reference {} Rubin, V., Ford, W.K., Thonnard, N 1978, ApJ 225, L107

\reference {} Sahu, K.C., 1994, Nature, 370, 275

\reference {} Sahu, K.C., 2001, in {\sl The Dark Universe: Matter, Energy, and Gravity},
STScI (in prep)

\reference {} Sahu, K.C., Sahu, M.S. 1998, ApJ 508, L147

\reference {} Siegel, M.H., Majewski, S.R., Reid, I.N. 2001, in. preparation.

\reference {} Schmidt, M. 1975 ApJ 202, 22

\reference {} Trimble, V. 1987, ARAA 25, 425

\reference {} Wielen, R. 1977, A\&A 60, 263

\reference {} Wilkinson, M.I., Evans, N.W. 1999, MNRAS 310, 645

\reference {} Zwicky, F. 1933, {\sl Helv. Phys. Acta} 6, 110

\end{references}
\end{document}